\documentclass[12pt,preprint,natbib]{emulateapj}

\def \HST{{\emph{HST}}}
\def \Spitzer{{\emph{Spitzer}}}
\def \tf {{24 $\mu$m}}

\begin{document}
\slugcomment{6/25/08}

\title{Morphologies of High Redshift, Dust Obscured Galaxies from Keck Laser Guide Star Adaptive Optics}
 
\author{J. Melbourne \altaffilmark{1}, V. Desai \altaffilmark{2}, Lee Armus \altaffilmark{2}, Arjun Dey \altaffilmark{3}, K. Brand \altaffilmark {4}, D. Thompson \altaffilmark{5}, B. T. Soifer \altaffilmark{1,2}, K. Matthews\altaffilmark{1}, B. T. Jannuzi \altaffilmark{3}, J. R. Houck \altaffilmark{6} }

\altaffiltext{1}{Caltech Optical Observatories, Division of Physics, Mathematics and Astronomy, Mail Stop 320-47, California Institute of Technology, Pasadena, CA 91125, jmel@caltech.edu, vandesai@gmail.com, bts@submm.caltech.edu, kym@caltech.edu}

\altaffiltext{2}{Spitzer Science Center, Mail Stop 314-6, California Institute of Technology, Pasadena, CA 91125, lee@ipac.caltech.edu, bts@ipac.caltech.edu}

\altaffiltext{3}{National Optical Astronomy Observatory, P.O. Box 26732, Tucson, AZ 85726-6732, dey@noao.edu, jannuzi@noao.edu}

\altaffiltext{4}{Space Telescope Science Institute, Baltimore, MD 21218, brand@stsci.edu}

\altaffiltext{5}{University of Arizona, LBT Observatory, 933 N. Cherry Ave. Tuscon, AZ 85721-0065, dthompson@as.arizona.edu}

\altaffiltext{6}{Astronomy Department, Cornell University, Ithaca, NY 14853, jrh13@cornell.edu}

\begin{abstract}
\Spitzer\ MIPS images in the Bo\"otes field of the NOAO Deep Wide-Field Survey have revealed a class of extremely dust obscured galaxy (DOG) at $z\sim2$.  The DOGs are defined by very red optical to mid-IR (observed-frame) colors, $R - [24\; \mu m] > 14$ mag, i.e. $f_{\nu} (24\; \mu m) / f_{\nu} (R) > 1000$.  They are Ultra-Luminous Infrared Galaxies with $L_{8-1000 \mu m}>10^{12}-10^{14} \;L_{\odot}$, but typically have very faint optical (rest-frame UV) fluxes.  We imaged three DOGs with the Keck Laser Guide Star Adaptive Optics (LGSAO) system, obtaining $\sim0.06\arcsec$ resolution in the $K'$-band.  One system was dominated by a point source, while the other two were clearly resolved.  Of the resolved sources, one can be modeled as a exponential disk system.  The other is consistent with a de Vaucouleurs profile typical of  elliptical galaxies.   The non-parametric measures of their concentration and asymmetry, show the DOGs to be both compact and smooth.  The AO images rule out double nuclei with separations of greater than $0.1\arcsec$ ($< 1$ kpc at $z=2$), making it unlikely that ongoing major mergers (mass ratios of 1/3 and greater) are triggering the high IR luminosities.  By contrast, high resolution images of $z\sim2$ SCUBA sources tend to show multiple components and a higher degree of asymmetry.  We compare near-IR morphologies of the DOGs with a set of $z=1$ luminous infrared galaxies (LIRGs; $L_{IR} \sim 10^{11} \; L_{\odot}$) imaged with Keck LGSAO by the Center for Adaptive Optics Treasury Survey.  The DOGs in our sample have significantly smaller effective radii, $\sim1/4$ the size of the $z=1$ LIRGs, and tend towards higher concentrations.  The small sizes and high concentrations may help explain the globally obscured rest-frame blue-to-UV emission of the DOGs.   
\end{abstract}


\section{Introduction}
The \Spitzer\ Space Telescope \citep{Werner04} has revealed large numbers of luminous and ultra-luminous infrared galaxies (LIRGs with $L_{IR} \sim 10^{11-12} \; L_{\odot}$, and ULIRGs with $L_{IR} > 10^{12} \; L_{\odot}$) in the distant universe \citep[e.g.,][]{Bell05, LeFloch05}.  Dust heating by star formation and active galactic nuclei (AGN) within the LIRGs and ULIRGs result in large mid-infrared fluxes detectable to high redshift by the \Spitzer\ Multiband Imaging Photometer \citep[MIPS;][]{Rieke04}.  Because much of the energy in these systems has been reprocessed by dust, optical (rest-frame UV) studies have underestimated their bolometric luminosities \citep{Bell05}.  At the extreme are sources that are highly obscured at observed optical wavelengths. 

Recent MIPS \tf\ images in the Bo\"otes field of the NOAO Deep Wide-Field Survey \citep[NDWFS;][]{JannuziDey99}  have revealed a class of extremely dust obscured galaxy \citep[DOG;][]{Houck05, Weedman06,Brand07}.  These systems were selected to have \tf\ flux densities in excess of 0.3 mJy, and optical to IR colors redder than $R-[24] >14$ mag, roughly a magnitude redder than Arp 220 at any redshift \citep{Dey08}.    Spectroscopic redshift confirmations for 86 DOGs were obtained with a combination of optical spectroscopy \citep[Keck DEIMOS and LRIS;][]{Desai07},  near-IR spectroscopy \citep[Keck NIRSPEC;][]{Brand07}, and mid-IR spectroscopy \citep[\Spitzer\ IRS;][Higdon et al., in preparation]{Houck05, Weedman06}.  The redshift distribution of these DOGs is Gaussian, with a mean of $z=1.99$, a sigma of $\sigma_z =0.5$, and a redshift range of $z=0.8 - 3.2$ \citep{Dey08}.

The rest-frame UV to mid-IR spectral energy distributions (SEDs) of DOGs range from power laws, rising into the mid-IR, to those that show a ``bump'' at rest wavelength $1.6 \; \mu m$ (observed in \Spitzer\ IRAC 3 - 8 $\mu m$ bands; Desai et al. in preparation, Dey et al. 2008).  A possible explanation for these two classes is that the power law sources are AGN dominated, whereas the SEDs of the ``bump'' sources are primarily powered by star formation.  For instance,  spectroscopic $H\alpha$ observations of 10 power law dominated, dust obscured galaxies in Bo\"otes showed that all 10 harbor an AGN \citep{Brand07}.  In reality, most DOGs are likely to contain some combination of both vigorous star formation and AGN activity \citep{Fiore08}.  

While spectra and SEDs may reveal the primary power sources for the DOGs, they do not necessarily reveal the triggering mechanism.  Are the DOGs undergoing violent mergers that drive both central star formation and AGN?  Alternatively, are the DOGs isolated massive galaxies seen at the time of their first collapse?  Are there other processes contributing such as bar formation or minor mergers?  Deep, spatially resolved imaging of the DOGs should help distinguish between some of these scenarios.  For instance, recent merging activity may be revealed by multiple nuclei or tidal tails \citep[e.g.,][]{Melbourneetal05}.  Smooth compact systems may be more readily explained by a primordial collapse model \citep[e.g.,][]{Zirm07}. Imaging may also be useful for identifying strong central point sources suggesting AGN activity.   In order to observe the morphology of the underlying stellar population, rest-frame optical to near-IR imaging is preferable to imaging at shorter wavelengths.  It is less affected by ongoing star formation and dust obscuration compared with rest-frame UV imaging.  Rest-frame UV, however, is valuable for identifying sites of recent star formation and unobscured AGN, which may or may not track the underlying stellar distribution.

We have undertaken two efforts to obtain resolved images of the DOGs.  This paper reports on high spatial resolution ($0.06\arcsec$) Keck laser guide star (LGS) adaptive optics (AO) $K'$-band imaging (rest-frame optical to near-IR) of three DOGs.  While this sample is small, it is the first and only set of high resolution $K$-band images available for the DOGs.  Bussman et al. (in preparation) will report on \emph{Hubble Space Telescope}\ (\HST) imaging of an additional 30 DOGs.  The Bussman et al. data include ACS  and WFPC2 $I$-band imaging which probe rest-frame blue to UV, and which will be valuable for detecting the sites of ongoing star formation.  The Bussman et al. data also include NICMOS $H$-band imaging (rest-frame optical), which will be more sensitive to the underlying stellar population.  In comparison to NICMOS, the Keck AO data presented in this paper probe longer wavelengths, and have roughly a factor of 3 higher spatial resolution, providing an unprecedented view of the stellar distributions in the cores of the DOGs.     %Unfortunately, the DOGs in the AO sample do not currently have spectroscopic redshifts.  However, as is shown in \citet{Dey07} the DOG selection criteria strongly selects for $z>1$ dust obscured galaxies.             

Section 2 describes the sample selection, the observations, and the data reduction.  Section 3 describes the morphologies of the DOGs and provides measurements of their structural parameters.  A discussion and comparison of the DOGs with two additional samples of high redshift dusty galaxies is given in Section 4.   Throughout, we report Vega magnitudes and assume a $\Lambda$ cold-dark-matter cosmology: a flat universe with $H_0= 70$ km/s/Mpc, $\Omega_m=0.3$, and $\Omega_{\Lambda}=0.7$.
 
\section{The Data}
Keck LGS AO observations of three DOGs were obtained on the night of U.T. 2007 May 21.  The AO system tracks and corrects wavefront distortions of astronomical sources as the light propagates through the Earth's turbulent atmosphere.  For a complete description of the Keck LGS AO system, see \citet{vanDam06}.  Despite the laser, the AO system still requires a nearby ($< 55\arcsec$) faint ($R<18$) natural guide star (``tip-tilt'' star) to correct for image motion. By correcting for the effects of turbulence, the AO system achieves diffraction limited spatial resolution. The Keck AO system has a $K$-band spatial resolution of $\sim0.06\arcsec$, comparable to \HST\ in the optical, and a factor of $\sim3$ higher resolution than \HST\ NICMOS in the near-IR.  The performance of the AO system can be summarized by the Strehl ratio \cite[e.g.,][]{Wizinowich00}. This is the ratio of the peak brightness of the observed point-spread-function (PSF) to the theoretical maximum peak of a perfect diffraction limited PSF.   Table \ref{tab:obs} gives a summary of the observations, including exposure times and AO performance as given by the full-width-half-maximum (FWHM) of the PSF and the Strehl ratio.  
\begin{figure}
\centering
\includegraphics[trim=43 0 0 0, scale=0.48]{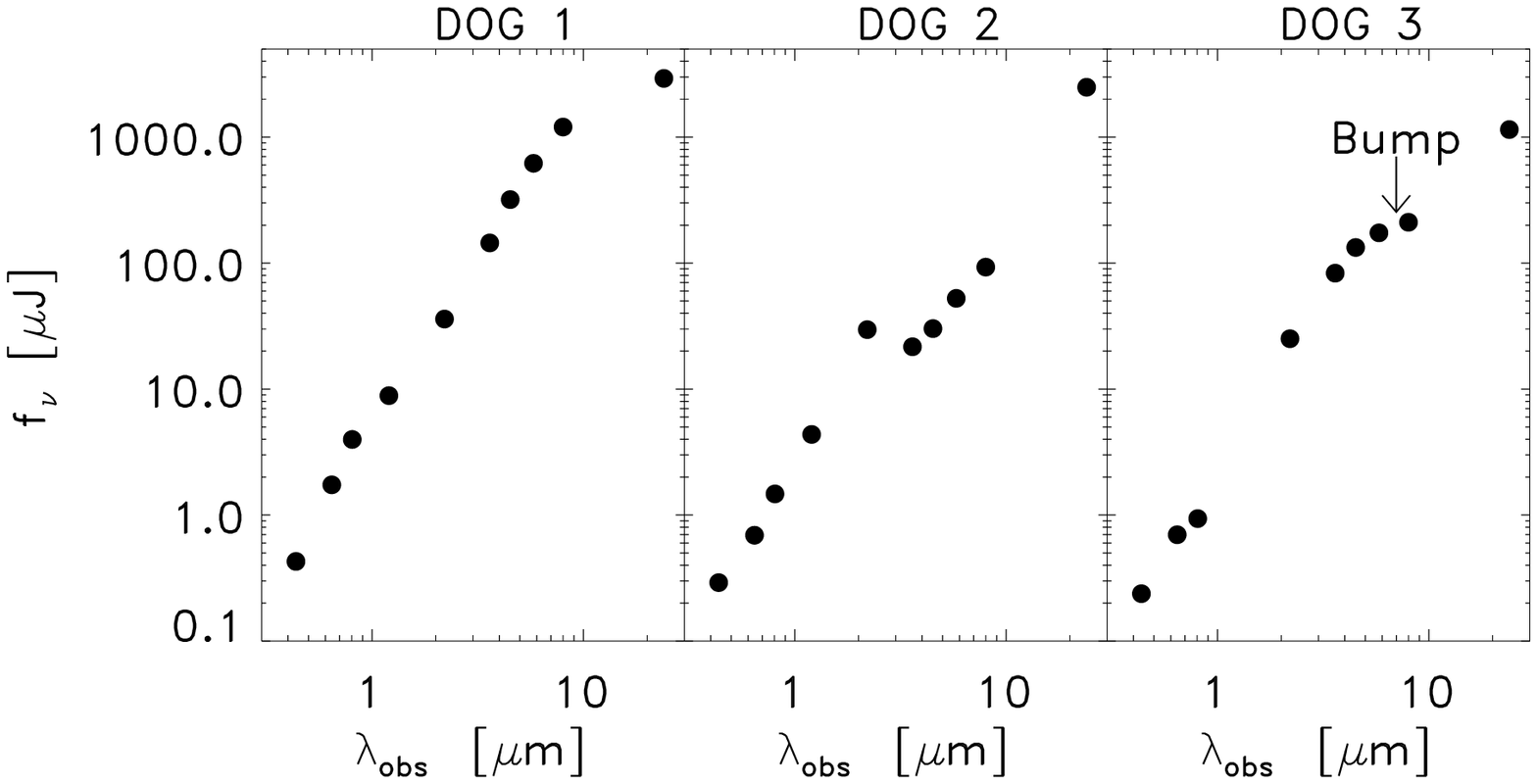}
\caption{\label{fig:sed} Spectral energy distributions (SEDs) of the three DOGs in our sample.  Photometric measurements were made from optical NDWFS $B_w,R,I$ ground based images \citep{JannuziDey99}, near-IR $J,Ks$ FLAMEX images \citep{Elston06}, the four \Spitzer\ IRAC channels \citep[3.6, 4.5, 5.8, and 8 $\mu m$][]{Eisenhardt04}, and the Spitzer mid-IR 24 $\mu m$ band \citep{Houck05}.  DOGs 1 and 2 show power-law SEDs, while the SED of DOG 3 appears to have a ``bump'' in the IRAC bands. Typical photometric uncertainties are roughly the size of the points.  }
\end{figure}

 \begin{figure}
\centering
\includegraphics[trim=20 0 0 0, scale=0.65]{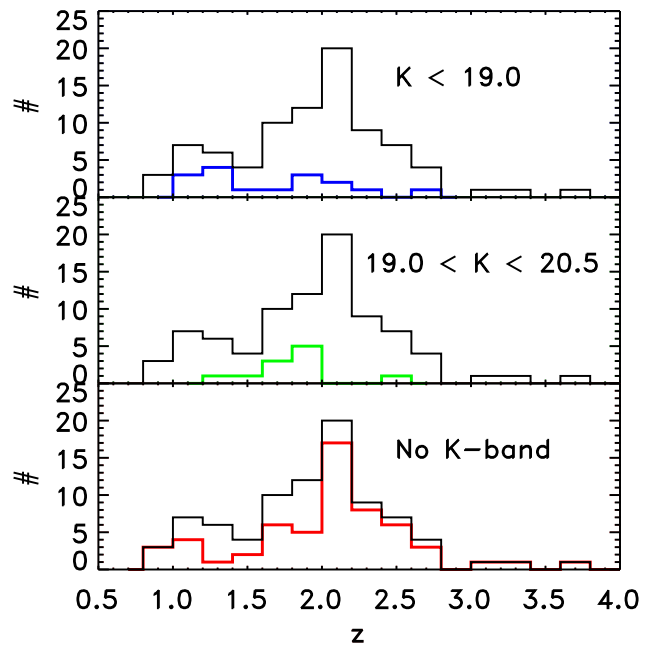}
\caption{\label{fig:zdist} The redshift distribution of the DOGs \citep[Black][]{Dey08}.  Also shown is the distribution for different $K$-band magnitude limits.  The $K$-band bright targets ($K < 19$, blue, top panel) favor lower redshifts ($z\sim1$) but still span the range of the full redshift distribution (Black).  In contrast, fainter $K$-band sources (green, middle panel) trace the larger distribution more closely.  A significant number of the DOGs with redshifts were either not observed at $K$, or were too faint to be detected (red, bottom panel).
}  
\end{figure}

\subsection{The Sample}
We identified a set of 2603 DOGs in the Bo\"otes field of the NDWFS \citep{Dey08}.  These galaxies were selected to have $R-[24]>14$, roughly $f_{\nu} (24 \;\mu m) / f_{\nu} (R) > 1000$, and \tf\ flux densities in excess of 0.3 mJy.  We matched these objects to the positions of suitable tip-tilt AO guide stars from the USNO-A2.0 catalogue \citep{Monet98}.   For this pilot program, we selected DOGs that were located within $30\arcsec$ of stars brighter than $R=15.5$, optimal conditions for achieving the highest level of AO correction. This provided us with a set of 55 potential targets.  

\begin{figure}
\centering
\includegraphics[trim=20 0 0 0, scale=0.65]{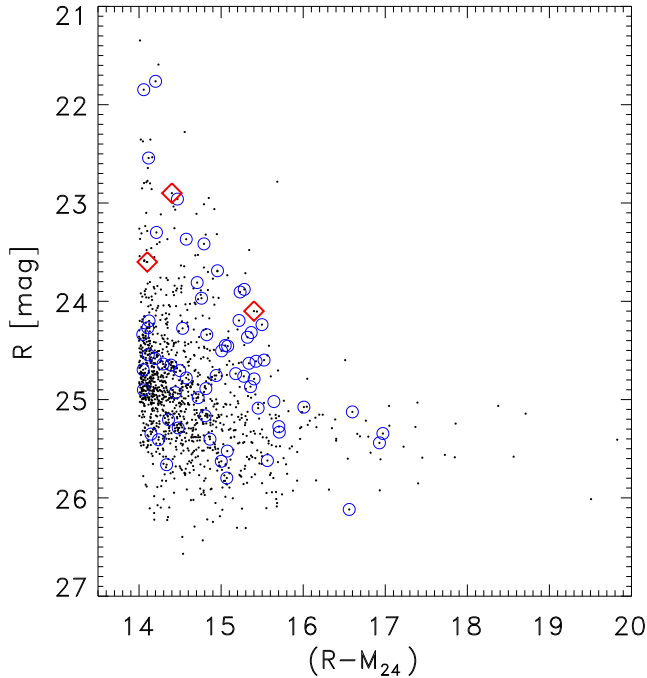}
\caption{\label{fig:aocomp} A color-magnitude diagram of the DOGs found in the NDWFS (black points).  DOGs with spectroscopic redshifts are circled in blue.  The AO sample is shown with red diamonds.  There does not appear to be anything unusual about the DOGs in the AO sample.  
}  
\end{figure}

We prioritized this sample by measured $Ks$-band magnitude \citep[from the FLAMEX survey of Bo\"otes;][]{Elston06}. For DOGs with no FLAMEX $Ks$ imaging, $Ks$ magnitudes were extrapolated from \Spitzer\ IRAC 3.6 $\mu m$ photometry.  Due to poor weather, we were only able to image three DOGs (denoted DOG 1, DOG 2, and DOG 3 for simplicity).  In order to ensure successful observations of a preliminary sample despite limited observing time, the three DOGs selected were among the brightest $Ks$-band sources in the sample ($Ks <19$). Note: while DOG 3 did not have a FLAMEX observation, its extrapolated $Ks$-magnitude was brighter than this limit.  Subsequent $Ks$ observations of DOG 3, made by our team with the Palomar Wide-field Infrared Camera \citep[WIRC;][]{Wilson03}, confirmed this.    With \tf\ fluxes $>1$ mJy, they also happened to be among the brightest \tf\ sources in our set of 55 possible targets.  The MIPS \tf\ fluxes and the FLAMEX and WIRC $Ks$-band magnitudes are given in Table \ref{tab:prop}.

A final consideration in the sample selection was given to the shape of the optical - IR SED of the DOGs.  
We chose our sample such that it would contain at least one of each SED type, power law, and ``bump''. Figure \ref{fig:sed} shows the SEDs of the three DOGs in our sample.  The ground-based optical data, $B_w$, $R$, and $I$-band filters, are from the NDWFS \citep{JannuziDey99}.  The ground-based near-IR data, $J$ and $Ks$-band filters, are from FLAMEX \citep{Elston06}.  The four space-based \Spitzer\ IRAC channels, 3.6, 4.5, 5.8, and 8 $\mu m$, were described by \citet{Eisenhardt04}, while the Spitzer mid-IR 24 $\mu m$ data were described by \citet{Houck05}.  Two of the galaxies selected, DOGs 1 and 2, show power law SEDs. DOG 3 is the most likely ``bump'' candidate.  While we succeeded in selecting at least one of each SED type, we caution that DOG 3 is not the best ``bump'' candidate in the larger DOG pool.  There are other DOGs in Bo\"otes that show a more prominent "bump" feature.  %Oddly enough as we will show in Section 3, DOG 3 is the only DOG in our AO sample that morphologically shows a strong central point source, usually indicative of an AGN.  This result was unexpected because the in-going hypothesis was that ``bump'' sources were powered by star formation, while power-law sources were AGN dominated.  This suggests that there may be a complex relationship between morphology and SED type, and that larger imaging samples are needed to understand it.

While DOG 2 appears to have a power law SED, the $Ks$-band data point is offset from the rest of the power law.  This data point is based on the FLAMEX $Ks$-band image, and is measured to within 10\%.  In addition, we checked the FLAMEX zeropoint with 2MASS images of the field and found good agreement.  It is possible that a strong emission line is contributing to the deviation of the $Ks$-band data point from the power law SED.  If the object is at $z\sim2$ then $H\alpha$ falls into the $Ks$ pass-band.  A very large, rest-frame $H\alpha$ equivalent width, on the order of 1000 \AA, could account for this offset.  This is about twice as large as the largest $H\alpha$ equivalent width found in the \citet{Brand07} sample of NIRSPEC DOG spectroscopy.  While this data point remains puzzling, the remainder of the analysis in this paper is independent of the total measured $K$-band flux. Instead we will be investigating how the flux is distributed spatially.  

Unfortunately, because of their proximity to bright AO guide stars, the three DOGs in the AO sample have not previously been targeted by our spectroscopic surveys.  While we do not know the exact redshifts of these galaxies, \citet{Dey08} has shown that the DOG selection criteria strongly selects for $z>1$ dust obscured galaxies.  Figure \ref{fig:zdist} shows the redshift distribution as given in \citet{Dey08}.  Also shown are the redshift distributions at different $K$-band flux levels.  While the $K$-band bright DOGs show a preference for the low-z end of the redshift distribution ($z\sim1$), they do span the range of the larger DOG sample.  As we primarily limit our analysis to morphology, a lack of redshift information does not prove critical to our analysis.  

Figure \ref{fig:aocomp} shows how the photometry of the AO sample compares to the larger DOG sample.  Aside from being at the brighter end of the distribution, there does not appear to be anything unusual about the three DOGs in the AO sample.  

 \begin{deluxetable*}{ccccccc}
\tabletypesize{\small}
\tablecaption{Keck Observation Summary \label{tab:obs}}
\tablehead{\colhead{object} & \colhead{MIPS Catalogue Name \tablenotemark{a}}   & \colhead{exptime } &  \colhead{FWHM}& \colhead{Strehl} & \multicolumn{2}{c}{Guide Star} \\ & & \colhead{[min]} & \colhead{of PSF [$\arcsec$]} &\colhead {estimate [\%]} &\colhead{$R$ [mag]} &\colhead{sep [$\arcsec$]}}
\startdata
DOG 1 &	SST24 J143234.9+333637 &  30 &  0.057&18 & 15.0 & 26.4\\
DOG 2 &  SST24 J142801.0+341525  &  30 & 0.053 & 24 &14.3 & 20.5\\
DOG 3 &  SST24 J142944.9+324332  &  9  & - & - &14.5 & 27\\
\enddata
\tablenotetext{a}{\citet{Houck05}}
\end{deluxetable*}

\subsection{AO Observations} 
AO observations were made in the $K'$-band with the NIRC2 camera.  In order to achieve the full Keck resolution, we observed in the narrow field mode which has an $0.01\arcsec$ pixel scale, and a $10\arcsec$ field of view.  Individual exposures were 180s and a dither was applied between exposures.  DOGs 1 and 2 were observed for 30 minutes each, while DOG 3 was observed for 9 minutes.  Observations of DOG 3 were cut short by a laser fault. 

\begin{figure*}
\centering
\includegraphics[scale=1, trim= 40 250 0 100 ]{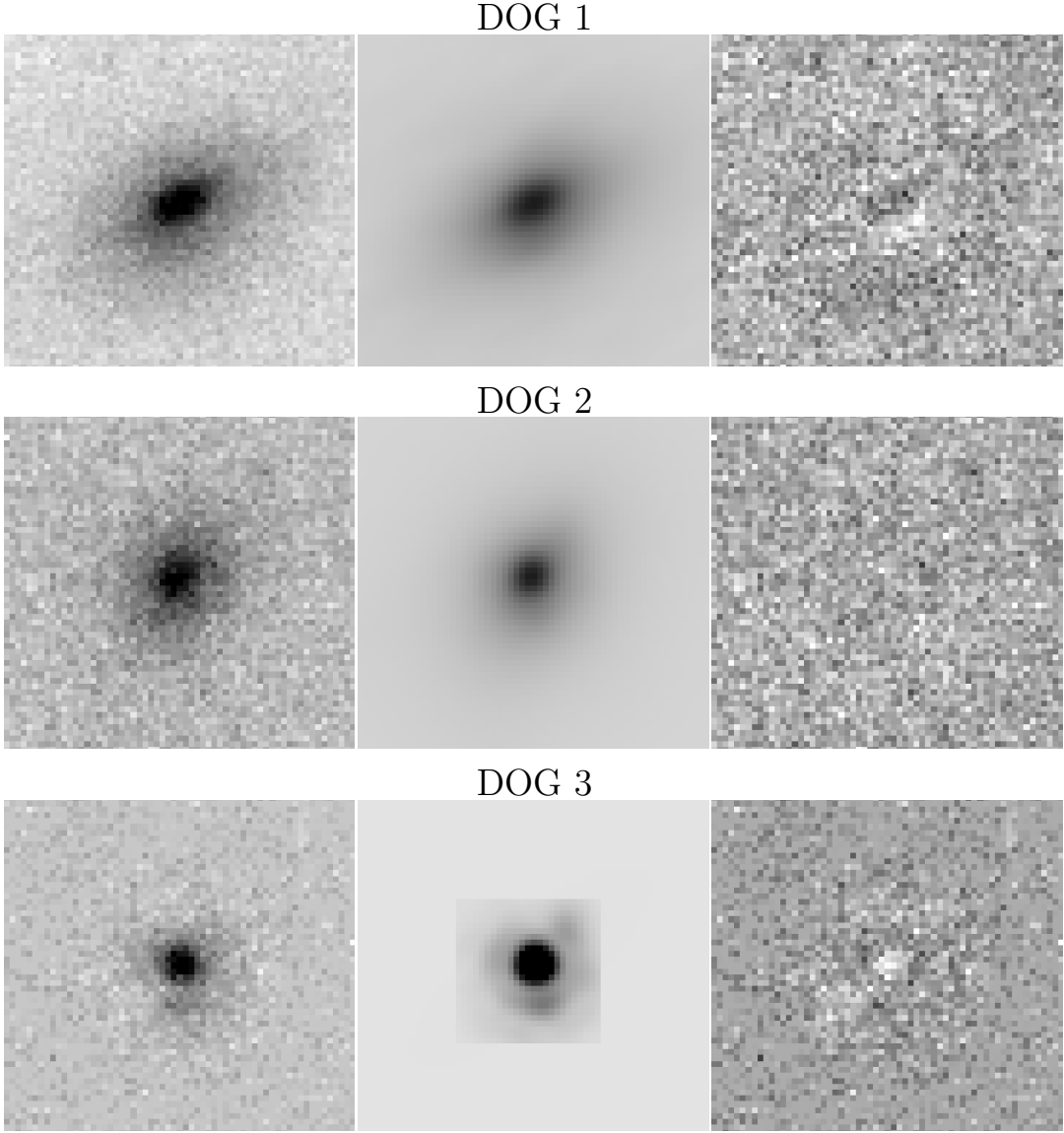}
\caption{\label{fig:dogs} $K'$ images of the DOGs 1 (top), 2 (middle), and 3 (bottom).  The left column gives the observation.  The middle column gives the best fit model of the light distribution.  The right column is the residual difference between the two (observation - model).  DOG 1 is best fit by an exponential disk with Sersic $n=1.5$.  DOG 2 is best fit by a  de Vaucouleurs profile (actual $n=3.5$).  DOG 3 appears to be dominated by a point-source. The images of the DOGs tend to be smooth showing little evidence for substructure, such as multiple nuclei. The faint substructures seen in the residual images, especially for DOGs 1 and 3, represent a small fraction of the total flux.  At the center of the galaxy, the RMS deviation of the residuals  is $ 5\%$  or less of the galaxy flux/pix.  The images are $\sim0.6 \arcsec$ on a side, with north up and east to the left.  }
\end{figure*}

In Section 3 we model the two-dimensional light profiles of our galaxies.  In order to do this accurately, we tracked the complicated AO PSF throughout the night.  The $10\arcsec \times 10\arcsec$ field surrounding DOG 1 contained a faint star suitable for tracking the real time variations of the AO PSF.  We used this PSF to model the spatial light profile of DOG 1.  In order to track the PSF of DOG 2 we obtained images of the tip-tilt star immediately following the science observations.  Because of the laser fault, we did not obtain a PSF specifically matched to DOG 3, but use both existing PSFs in the analysis of the DOG 3 light profile.  %As we will show in Section 3, DOG 3 is dominated by a strong central point source.  This conclusion holds regardless of whether we use the DOG 1 PSF or the DOG 2 PSF, when analyzing the morphology of DOG 3. Unfortunately, any underlying galaxy (after subtraction of the central point source) for DOG 3 is difficult to model because of low signal-to-noise ratio.

\subsection{Data Reduction}
The observations were reduced in a manner similar to \citet{MelbourneSN07}. Frames were corrected for flatfield variations with an average combined dome flat.  Frames were then sky-subtracted using a scaled clipped mean sky image, obtained from the actual data frames with sources masked out.  Images were aligned by centroiding on objects in the field, and then combined with a clipped mean. 

Because there was significant cirrus during the run,  zeropoints were set by measuring the magnitudes from the FLAMEX  \citep{Elston06}  and WIRC $Ks$-band images.  Because the $Ks$ filter (central wavelength $= 2.15 \; \mu m$, bandpass width $= 0.31 \; \mu m$) is only slightly narrower than $K'$ (central wavelength $= 2.12 \; \mu m$, bandpass width $= 0.35 \; \mu m$), the magnitude differences between $K'$ and $Ks$ should be less than 0.1 magnitudes.  As 0.1 magnitudes is roughly our photometric uncertainty, we choose to report all measured magnitudes in the $Ks$ filter. 

\section{Morphology and Photometry}

Figure \ref{fig:dogs} shows the AO images of the three DOGs in our sample (left panels).  Two of the three systems, DOGS 1 and 2, are resolved.  The third appears to be dominated by a point source, possibly the result of an AGN.  Interestingly, there is no significant substructure in these images.  The galaxy light profiles appear to be smooth, as opposed to comprised of multiple knots or nuclei.%, suggesting that these systems are not in the early stages of a major.  

We used the 2D galaxy profile fitting routine GALFIT (Peng et al. 2002) to model each system.  We provided GALFIT with the image of each galaxy, its associated PSF, and an initial guess for the best fitting galaxy model.  GALFIT then matched for total magnitude, effective radius, semi-minor to semi-major axis ratio, and position angle.  We first ran each system with a simple, single Sersic model \citep{Sersic68},
\begin{equation}
\Sigma(r)=\Sigma_e exp[-\kappa((\frac{r}{r_e})^{1/n}-1)],
\end{equation}
allowing the Sersic parameter, `$n$', and the effective radius, $r_e$, to float.  The best fitting single Sersic parameter for each system is reported in Table \ref{tab:prop}.  

Images of the GALFIT derived models are shown in the middle column of Figure \ref{fig:dogs}.  The right hand panels show the residuals after subtracting the model from the actual galaxy.  DOG 1 is best fit by an exponential disk (actual best fit Sersic is $n=1.5$) with an effective radius of $0.11 \arcsec$.    DOG 2 is best fit by a  de Vaucouleurs  profile (actual best fit Sersic is $n=3.5$) with effective radius $0.09 \arcsec$.  

While both DOGs 1 and 2 are resolved they have very small effective radii.  If they are at $z=1$, $0.09\arcsec$ translates to a physical size of $\sim0.72$ kpc.  Because of the geometry of the universe, apparent size, for a given physical size, remains roughly constant with redshift above $z=1$.  That means that even if DOGs 1 and 2 are at $z=2$, their physical sizes remain small.  At $z=2$, $0.09 \arcsec$ translates to $\sim0.75$ kpc.  These are very small sizes, more typical of today's blue compact dwarfs \citep[e.g.,][]{Corbin06} than today's large star forming ULIRGs.     
 
 \begin{deluxetable}{c|ccc}
\tabletypesize{\small}
\tablecaption{Photometric and Morphological Properties of the DOGs \label{tab:prop}}
\tablehead{\colhead{Property} & \colhead{DOG 1} & \colhead{DOG 2} & \colhead{DOG 3} }
\startdata
\tf\ flux \tablenotemark{a} [mJy]& 2.92 &2.49  & 1.15 \\
$K_s$ [mag] & 18.21 \tablenotemark{b} & 18.42 \tablenotemark{b} &18.60 \tablenotemark{c} \\
Morphology  \tablenotemark{d} & disk & elliptical & point source\\
Single Sersic Index, n \tablenotemark{e}  &  1.54 & 3.48 & 6 \\
Effective Radius \tablenotemark{f} [$\arcsec$]  & 0.11  & 0.09 & 0.02\\
 C \tablenotemark{g} & 3.3 & 3.8 & 3.8\\
 A \tablenotemark{h} & 0.08& 0.05 & 0.39\\
 
\enddata
\tablenotetext{a}{From \citet{Houck05}.}
\tablenotetext{b}{From FLAMEX \citep{Elston06}.}
\tablenotetext{c}{From WIRC.}
\tablenotetext{d}{A visual classifciation from the AO images.}
\tablenotetext{e}{As measured by GALFIT}
\tablenotetext{f}{As measured by GALFIT}
\tablenotetext{g}{Concentration}
\tablenotetext{h}{Asymmetry}

 \end{deluxetable}
 
DOG 3 is best fit by a point source of magnitude $Ks\sim18.6$.  The structure seen in the image of DOG 3 appears to be primarily the result of structure in the AO PSF rather than ``true'' structure in the galaxy.  This PSF structure is demonstrated in the GALFIT model of DOG 3 (middle column) which is effectively a scaled version of the observed PSF (in this example, we used images of the tip-tilt star for DOG 2 as the PSF).    

After subtracting the best fitting model galaxies from the actual DOG images, the residual images (right hand column of Figure \ref{fig:dogs}) show little additional substructure; the possible exception being DOG 1.  Any remaining flux within the residuals is small,  with RMS deviations less than 5\% of the flux/pix at the centers of each DOG.   DOG 1 is the only object of the three suggestive of a secondary central component.   We used GALFIT to simultaneously fit DOG 1 for both a disk and a bulge.  The fit for the central component is actually disk-like with a Sersic index of $n=0.3$, similar to the truncated central disk in NGC 2787 \citep{Peng02}.  Although GALFIT produced a fit for this second component the light within it was roughly a factor of 100 less than in the primary component, and the improvement over a single Sersic model was negligible.  

In order to compare the morphologies of the DOGs with other samples of high redshift dusty galaxies, we also measured two non-parametric morphological indexes, concentration (C), and asymmetry (A).  We used the C and A definitions given in \citet{Conselice03}.  $C = 5log(r_{80}/r_{20})$, where $r_{80}$ is the radius that contains 80\% of the light, and $r_{20}$ is the radius containing 20\% of the light.   We measured the concentration index with circular aperture photometry and a curve of growth technique.  To provide a reference, we first measured the concentration of the AO PSF, which has $C\sim5$.   While all three DOGs show high concentrations, $4>C>3$, typical of early-type disks and elliptical  galaxies in the local universe \citep{Conselice03}, they are less concentrated than the AO PSF.  This is true even for the point-like DOG 3.

\begin{figure}
\centering
\includegraphics[trim=30 0 0 0,scale=0.65]{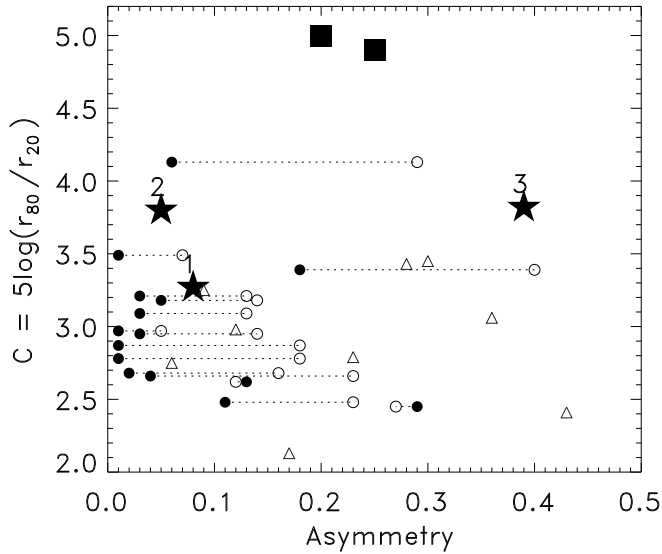}
\caption{\label{fig:cas} Asymmetry (A) and concentration (C) measurements of dusty galaxies.  Filled symbols are asymmetry measurements made in the $K'$-band.  Open symbols are asymmetry measurements made in $z$-band.  Measurements of the DOGs are shown as stars, and are labeled.  A and C measurements for a sample of $z\sim2$ SCUBA sources measured in \HST\ ACS $z$-band images are shown as triangles.  A second comparison sample of $z\sim1$ LIRGs observed in both the $K'$ and $z$-band are shown as circles.  Lines are drawn between the two A measurements of each LIRG, with concentration measures made in the $K'$-band. The C and A measures for the two AO PSF stars are also shown (squares).   The spatially resolved DOGs show low asymmetry compared with the SCUBA sources.  The LIRGs show similarly low asymmetry in the $K'$ band images.  Interestingly, the LIRGs show much higher asymmetry in the $z$-band (rest-frame blue-UV).   The DOGs tend towards higher concentrations than the comparison samples.}
\end{figure}

Asymmetry is given by the normalized residuals after subtracting a 180-degree, rotated image of the galaxy from itself.  For a complete description of the asymmetry parameter, see \citet{Conselice00}.  Because asymmetry measures are dependent on identifying the galaxy center, we take the \citet{Conselice03} suggestion to use the minimum A after using a grid of rotation centers.  Not  surprisingly, the smooth images of DOGs 1 and 2 show small asymmetries, $A < 0.1$,  typical of non-interacting galaxies in the local universe \citep{Conselice03}.  DOG 3, which is point-source dominated, has a higher A measure than the other two, A=0.39.  Some portion of the asymmetry measured in DOG 3 can be attributed to structure within the AO PSF.  This structure can be seen in the GALFIT model image for DOG 3 (Figure \ref{fig:dogs}).  The asymmetry measures of the AO PSFs are $A\sim0.23$.
The non-parametric morphology measurements are summarized in Table \ref{tab:prop}.  

The uncertainties in these morphological measures are comprised of two distinct components, photometric uncertainties and PSF uncertainties.  The high signal-to-noise ratio images of DOGs 1 and 2 translate into low photometric uncertainties.  Because DOG 3 was observed for significantly less time, photometric uncertainties are more significant.  To quantify the affect of photometric uncertainties on our morphological measurements we have populated the AO images with model galaxies using the IRAF ARTDATA package.  These models were created to have similar morphological properties as the actual DOGs, but are placed at 10 different locations in the AO frames.  For DOGs 1 and 2 morphological measures of the 10 models show small dispersion, typically 5\% or less for both the GALFIT and C and A measures. For DOG 3 the dispersion reaches 20\%.  

To quantify the contribution of PSF uncertainty to our overall error budget, we run the GALFIT routine with both the best guess PSF, and with a second PSF. For DOG 1, we use the PSF for DOG 2 as the secondary PSF.  For DOG 2 we use the PSF for DOG 1.  Since no PSF measurement was made for DOG 3 we use PSFs 1 and 2 and compare.  For DOGs 1 and 2, the change of PSF has a significant affect on the measured Galfit parameters, changing them on the order of 20\%.  For DOG 3 the change in the measured Galfit parameters from PSF 1 to PSF 2 is only 10\%, less than the uncertainty from photometric noise.  While we do not know the true uncertainties in the PSFs, these are good upper bounds on PSF affects on the Galfit morphological measures. 

\begin{figure}
\centering
\includegraphics[scale=0.55, trim=0 0 0 20]{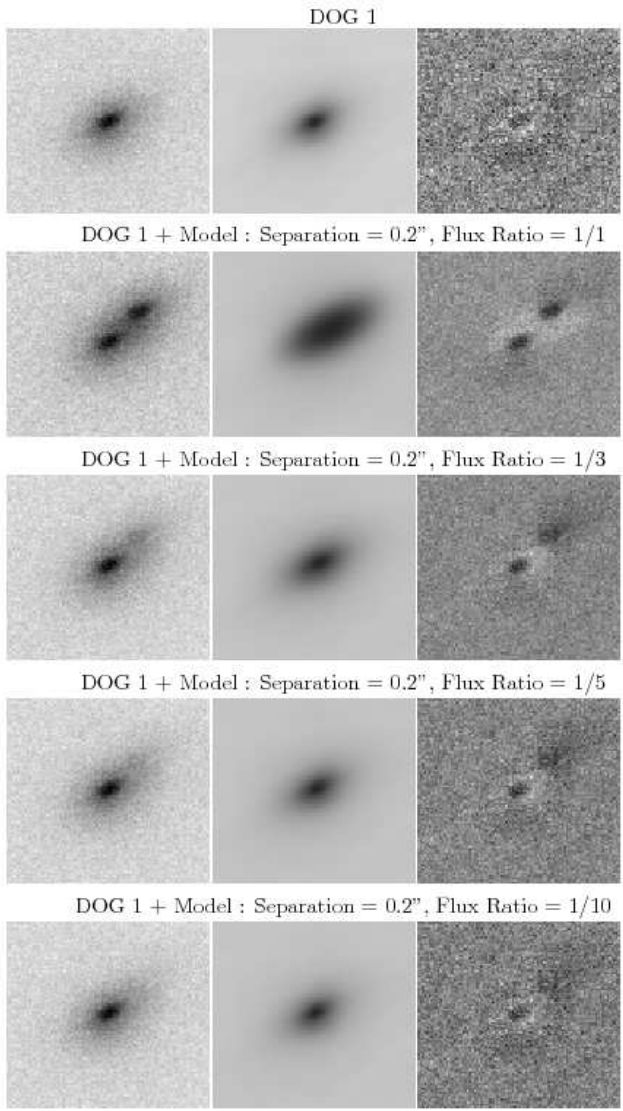}
\caption{\label{fig:model_sep20} Simulated AO images of mergers.  Images are $1\arcsec$ on a side. The mergers are based on the AO image of DOG 1. Two copies of the DOG 1 image are added together after shifting spatially and scaling to different flux ratios.  The left column shows the image of the  merger.  The middle column is the best fitting single Sersic model, as given by GALFIT.  The right hand column gives the residual difference between the two.  For comparison, the top row shows the image of DOG 1 with no merger added.  Successive rows show mergers with declining flux ratios, 1/1, 1/3, 1/5, and 1/10.  The merger components are separated by $0.2 \arcsec$ or 1.6 kpc at $z=2$.  This is well below the seeing limited resolution of Keck and at or below the spatial resolution of \HST\ NICMOS in the near-IR.   At these separations we can clearly identify mergers with flux ratios of 1/5 and larger.      
}
\end{figure}

\begin{figure}
\centering
\includegraphics[trim=30 0 0 0,scale=0.52]{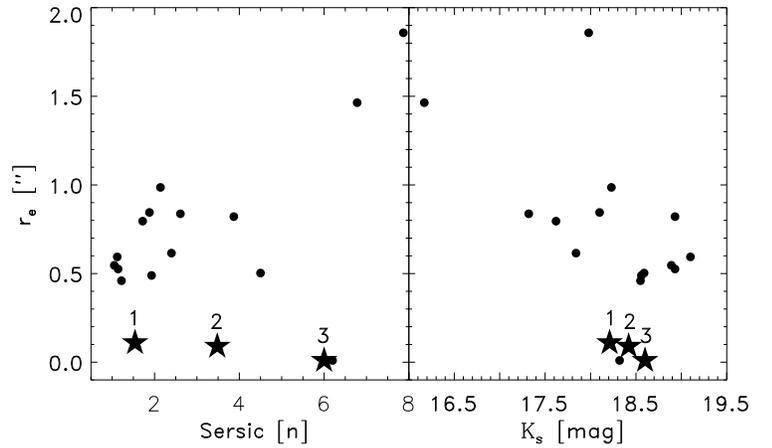}
\caption{\label{fig:sersic} The best fit single Sersic models of the DOGs (stars).  Also shown are the comparison sample $z\sim1$ LIRGs (circles).  The left panel shows effective radius as a function of Sersic index.  The right panel shows effective radius as a function of $Ks$ magnitude.  The two resolved DOGs show significantly smaller (roughly by a factor of four) effective radii compared with LIRGs of similar Sersic index or magnitude.  If we translate these apparent sizes into physical sizes using the range of possible DOG redshifts $z=1 - 3$, then the physical sizes of the DOGs are also about a factor of four smaller than the LIRGs for the full range of redshifts.  This is because apparent sizes are relatively fixed for a given physical size above redshift 1. The DOGs are fundamentally different from $z=1$ LIRGs.}
\end{figure}

\section{Discussion}

To place the three DOGs in our sample into context, we compare their morphological parameters with two additional samples of intermediate to high redshift, dusty galaxies.  The first is a sample of of $z=2$ SCUBA sources \citep[sub-mm sources,][]{Pope05} observed in the Great Observatories Origins Deep Survey  north field \citep[GOODS-N;][]{Giavalisco04}.  Although SCUBA sources have a different selection criteria, they are also very dusty.  The sub-mm emission is thought to arise from cool dust in very strongly star-bursting galaxies. The second is a sample of $z\sim1$ LIRGs from GOODS-S observed in the $K'$-band with Keck AO by the Center for Adaptive Optics Treasury Survey \citep[CATS;][]{Melbourne08}.  These two samples where chosen because they contain dusty galaxies, and because they span the redshift range of the DOGs.  

Figure \ref{fig:cas} is a plot of asymmetry vs. concentration for the DOGs in our sample (stars).  Also plotted are the GOODS-N SCUBA sources (triangles) and the GOODS-S LIRGs (circles).  The C and A measures for the two AO PSF stars are shown as well (squares). For the SCUBA sources, we only plot C and A measurements of galaxies that were reported to have high signal-to-noise ratio \citep{Pope05}.  The A and C measurements for SCUBA sources were made in the reddest of the GOODS \HST\ ACS bands, the $z$-band.  These images are  significantly bluer than our $K'$-band images of the DOGs.  However, \citet{Pope05} suggest that there is little change in the A and C measurements through to the near-IR $H$-band (based on a small subset for which NICMOS $H$-band is available).  They propose that the high asymmetry in the SCUBA sample indicates significant recent merging.  

Also plotted on Figure \ref{fig:cas} are the A and C measurements of a set of $z\sim1$ LIRGs from CATS \citep{Melbourne08}.  For the LIRGs, both $K'$- band (filled circles), and $z$-band (open symbols) asymmetries are shown.  The $K'$-band asymmetries of the LIRGs  are low in comparison with the SCUBA sources.  This was anticipated because \citet{Melbourne08} demonstrated that the morphologies of these LIRGs are primarily disk-like and unlikely to have had a recent major merger.  Interestingly, if we measure the asymmetries of the LIRGs in bluer bands, for instance the $z$-band (rest-frame optical to UV),  we find that they scatter towards higher asymmetries, similar to the SCUBA sources.   The LIRGs are more asymmetric in the bluer bands because of widespread star formation and dust obscuration, rather than merging. This suggests that interpretation of high asymmetry is difficult, especially when looking at rest-frame blue to UV images.   

The two resolved DOGs in our sample have, on average, lower asymmetry than the SCUBA sources.  Their asymmetry measures are similar to the $K'$-band asymmetries of the LIRGs. This may be evidence against recent merging within the DOGs.    In order to quantify this we generate model merger systems and analyze them with GALFIT.  The models are based on the image of DOG 1.  We add two copies of the DOG 1 image together after shifting spatially and scaling to different flux ratios.  Figure \ref{fig:model_sep20} shows model mergers generated with an $0.2\arcsec$ separation. This separation is well below the resolution of seeing limited images and at or below the resolution of \HST\ NICMOS in the near-IR. As with Figure \ref{fig:dogs}, the left column shows the merger images.  The middle column shows the best fitting single Sersic model for each merger, provided by GALFIT.  The right hand column gives the residual difference between the two.  For comparison, the top row shows the image of DOG 1 without any merger added, and a single Sersic model subtracted off.  Successive rows show mergers with decreasing flux ratios, of 1/1, 1/3, 1/5, and 1/10.  Visually, at these separations, 1/5 ratio mergers are easily detected in the residual images, but 1/10 ratio mergers are not obvious.  When we cut the separation in half to $0.1\arcsec$, 1/5 ratio mergers become more difficult  to identify, but 1/3 ratio mergers are still relatively easy to detect.  Thus the AO images rule out 1/5 ratio mergers at $0.2\arcsec$ separations, and 1/3 ratio mergers at $0.1\arcsec$ separations.  If the DOGs in our sample are at $z=2$, $0.1\arcsec$ translates to a physical separation of 0.8 kpc.  This is a strong constraint against ongoing major mergers as the trigger for the DOGs' IR energy.  Note, we can not rule out late stage mergers in which the nuclei have already merged completely.   

%Preliminary results show higher asymmetries for the DOGs at shorter wavelengths (Bussman et al., in preparation).  Our experience with the LIRGs suggests that we should be cautious before assigning a merger trigger to high asymmetry DOGs if the measurements are made in the rest-frame UV.

While the LIRGs and the resolved DOGs in our sample have similar $K'$-band asymmetries, the DOGs have, on average, higher concentrations.  The DOGs also tend to be more highly concentrated than the SCUBA sources.  We will discuss possible implications for the high concentration below, but first we compare the physical sizes of the DOGs and the $z=1$ LIRGs.

In addition to measuring non-parametric morphologies, we have measured 2D galaxy profiles for the DOGs.  As we showed in the previous section, the three DOGs in our sample show a  range of morphologies from disk-like (DOG 1), to elliptical-like (DOG 2), to point-like (DOG 3).  Figure \ref{fig:sersic} shows the best fit single Sersic parameters for the DOGs in our sample (stars).  We make the same measurements on the AO data for the $z\sim1$ LIRGs (circles).  Apparent sizes for the two resolved DOGs are a factor of 4 - 5 smaller than for LIRGs with similar Sersic indices or $K$-band magnitudes.  We do not know the redshifts for the three DOGs in our sample, therefore we cannot directly compare physical radii. We do, however, know that DOGs are typically at $z\sim2$ and span a range from $z\sim1$ to $3$.  If the DOGs in our sample happen to be at the same redshift as the LIRGs ($z\sim1$),  then their physical radii are smaller than the LIRGs by about factor of four.  Above $z=1$, apparent size remains relatively constant with redshift for a fixed physical size.  This means that even if the DOGs are at $z=3$ they still have sizes about a factor of four smaller than the LIRGs of $z=1$.    

In order to test whether the small sizes observed for the DOGs are the result of an observational bias against measuring objects with extended radii, we embedded model galaxies into the DOG frames.  These galaxies were designed to have the same $K$-band fluxes as the DOGs, but with radii similar to the LIRGs.  While these model objects were lower surface brightness than the actual DOGs, we were able to accurately measure their sizes and luminosities with GALFIT to within better than 10\%.   This argues that the DOGs really do have small effective radii. It is a further indication that the DOGs are not the same type of objects as the $z=1$ LIRGs. 

So what are the DOGs?  The DOGs have very red rest-frame UV to IR luminosities, a magnitude redder than Arp 220.  This suggests that, not only do they contain significant amounts of dust, but the dust may be ubiquitous, blocking the majority of UV emitting sources.  While our sample is currently very small, it may turn out that in order to be selected as a DOG, a galaxy must have a high concentration and small physical size.  Otherwise the likelihood of observing some UV light, unobscured by dust, increases.  This hypothesis will be interesting to explore with larger samples.

Another key piece of information for explaining the nature of the DOGs may be their SEDs.  At least two of the DOGs show power-law SEDs suggestive of AGN activity.  Oddly enough those two DOGs do not show a significant central point source in the AO image.  Both of these sources, however, have more highly concentrated light profiles than the LIRGs.  This may mean that they contain a dust enshrouded AGN, increasing their concentration, even though the AGN is not obvious in the images.  Based on a stacking analysis of very deep x-ray data in the Chandra Deep Field South \citet{Fiore08} show that DOGs in GOODS-S are likely to contain dust obscured AGN.  DOG 3 does show point-like morphology suggesting that it too may contain a strong central AGN.  Surprisingly, DOG 3 shows the least power-law like SED.  The ``bump'' in the SED of DOG 3 may be indicating that there is also strong star formation in this galaxy, and while the morphology is point like, the  C and A measures for DOG 3 suggest that it is not a pure point source galaxy.  Clearly we need to increase the sample size in order to better understand the relationship between morphology and SED type in the DOG population.  

In addition to morphology, spatially resolved spectroscopy and kinematics may help place further constraints on AGN activity or mergers in the DOG sample. It may be possible to obtain such data with AO-fed integral field unit (IFU) spectrographs on large telescopes \citep[e.g.,][]{Wright07}.  Spatially resolved, rest-frame optical spectroscopy targeting $H\alpha$ and [N II]  can reveal the presence and location of AGN \citep[e.g.,][]{Brand07}.  In addition, IFU spectroscopy can provide a measure of the kinematics within a galaxy.  While one of our systems has a disk-like light profile, we do not know if it is a normal rotating disk.  Spatially resolved spectroscopy might reveal that this disk-like system is actually kinematically disturbed in some way.    

We have demonstrated the power of high spatial resolution AO imaging for measuring the morphologies of dust obscured galaxies beyond $z=1$.  We plan to continue imaging the DOGs with Keck AO, and hope to add kinematic information with AO-fed integral field spectrographs.  These future observations may untangle the nature of the DOGs and help explain the formation of today's massive galaxies.       

\acknowledgments
The adaptive optics data presented herein were obtained at the Keck Observatory, which is operated as a scientific partnership among the Caltech, UC, and NASA.  The authors wish to recognize and acknowledge the very significant cultural role and reverence that the summit of Mauna Kea has always had within the indigenous Hawaiian community. The laser guide star adaptive optics system was funded by the W. M. Keck Foundation. The artificial laser guide star system was developed and integrated in a partnership between the Lawrence Livermore National Labs (LLNL) and the W. M. Keck Observatory. The laser was integrated at Keck with the help of Curtis Brown and Pamela Danforth. The NIRC2 near-infrared camera was developed by Caltech and UCLA  (P.I. Keith Matthews).   
This work is based in part on observations made with the \Spitzer\ Space Telescope, which 
is operated by the Jet Propulsion Laboratory, California Institute of Technology under NASA 
contract 1407.  The project was funded in part by the \Spitzer\ Science Center.  
We are especially grateful to the IRAC Shallow Survey team (P.I.  Peter Eisenhardt) for providing observations in the Bo\"otes field.  We want to thank Dan Weedman, Steven Willner, and Daniel Stern for their helpful comments on the manuscript.
We are grateful to the expert assistance of the staff of Kitt Peak National 
Observatory where the ground-based Bo\"otes field observations of the NDWFS were obtained. The authors 
thank NOAO for supporting the NOAO Deep Wide-Field Survey. The research activities of AD  and BTJ are supported by NOAO, which is operated by the Association of Universities for  Research in Astronomy (AURA) under a cooperative agreement with the National Science Foundation.

\bibliographystyle{apj}
\bibliography{/Users/jmel/bib/bigbib2}

\begin{thebibliography}{29}
\expandafter\ifx\csname natexlab\endcsname\relax\def\natexlab#1{#1}\fi

\bibitem[{{Bell} {et~al.}(2005){Bell}, {Papovich}, {Wolf}, {Le Floc'h},
  {Caldwell}, {Barden}, {Egami}, {McIntosh}, {Meisenheimer},
  {P{\'e}rez-Gonz{\'a}lez}, {Rieke}, {Rieke}, {Rigby}, \& {Rix}}]{Bell05}
{Bell}, E.~F., {Papovich}, C., {Wolf}, C., {Le Floc'h}, E., {Caldwell},
  J.~A.~R., {Barden}, M., {Egami}, E., {McIntosh}, D.~H., {Meisenheimer}, K.,
  {P{\'e}rez-Gonz{\'a}lez}, P.~G., {Rieke}, G.~H., {Rieke}, M.~J., {Rigby},
  J.~R., \& {Rix}, H.-W. 2005, \apj, 625, 23

\bibitem[{{Brand} {et~al.}(2007){Brand}, {Dey}, {Desai}, {Soifer}, {Bian},
  {Armus}, {Brown}, {Le Floc'h}, {Higdon}, {Houck}, {Jannuzi}, \&
  {Weedman}}]{Brand07}
{Brand}, K., {Dey}, A., {Desai}, V., {Soifer}, B.~T., {Bian}, C., {Armus}, L.,
  {Brown}, M.~J.~I., {Le Floc'h}, E., {Higdon}, S.~J., {Houck}, J.~R.,
  {Jannuzi}, B.~T., \& {Weedman}, D.~W. 2007, \apj, 663, 204

\bibitem[{{Conselice}(2003)}]{Conselice03}
{Conselice}, C.~J. 2003, \apjs, 147, 1

\bibitem[{{Conselice} {et~al.}(2000){Conselice}, {Bershady}, \&
  {Jangren}}]{Conselice00}
{Conselice}, C.~J., {Bershady}, M.~A., \& {Jangren}, A. 2000, \apj, 529, 886

\bibitem[{{Corbin} {et~al.}(2006){Corbin}, {Vacca}, {Cid Fernandes}, {Hibbard},
  {Somerville}, \& {Windhorst}}]{Corbin06}
{Corbin}, M.~R., {Vacca}, W.~D., {Cid Fernandes}, R., {Hibbard}, J.~E.,
  {Somerville}, R.~S., \& {Windhorst}, R.~A. 2006, \apj, 651, 861

\bibitem[{{Desai et al.}(2007)}]{Desai07}
{Desai et al.} 2007, \apj, Submitted

\bibitem[{{Dey} {et~al.}(2008){Dey}, {Soifer}, {Desai}, {Brand}, {Le Floc'h},
  {Brown}, {Jannuzi}, {Armus}, {Bussmann}, {Brodwin}, {Bian}, {Eisenhardt},
  {Higdon}, {Weedman}, \& {Willner}}]{Dey08}
{Dey}, A., {Soifer}, B.~T., {Desai}, V., {Brand}, K., {Le Floc'h}, E., {Brown},
  M.~J.~I., {Jannuzi}, B.~T., {Armus}, L., {Bussmann}, S., {Brodwin}, M.,
  {Bian}, C., {Eisenhardt}, P., {Higdon}, S.~J., {Weedman}, D., \& {Willner},
  S.~P. 2008, \apj, 677, 943

\bibitem[{{Eisenhardt} {et~al.}(2004){Eisenhardt}, {Stern}, {Brodwin}, {Fazio},
  {Rieke}, {Rieke}, {Werner}, {Wright}, {Allen}, {Arendt}, {Ashby}, {Barmby},
  {Forrest}, {Hora}, {Huang}, {Huchra}, {Pahre}, {Pipher}, {Reach}, {Smith},
  {Stauffer}, {Wang}, {Willner}, {Brown}, {Dey}, {Jannuzi}, \&
  {Tiede}}]{Eisenhardt04}
{Eisenhardt}, P.~R., {Stern}, D., {Brodwin}, M., {Fazio}, G.~G., {Rieke},
  G.~H., {Rieke}, M.~J., {Werner}, M.~W., {Wright}, E.~L., {Allen}, L.~E.,
  {Arendt}, R.~G., {Ashby}, M.~L.~N., {Barmby}, P., {Forrest}, W.~J., {Hora},
  J.~L., {Huang}, J.-S., {Huchra}, J., {Pahre}, M.~A., {Pipher}, J.~L.,
  {Reach}, W.~T., {Smith}, H.~A., {Stauffer}, J.~R., {Wang}, Z., {Willner},
  S.~P., {Brown}, M.~J.~I., {Dey}, A., {Jannuzi}, B.~T., \& {Tiede}, G.~P.
  2004, \apjs, 154, 48

\bibitem[{{Elston} {et~al.}(2006){Elston}, {Gonzalez}, {McKenzie}, {Brodwin},
  {Brown}, {Cardona}, {Dey}, {Dickinson}, {Eisenhardt}, {Jannuzi}, {Lin},
  {Mohr}, {Raines}, {Stanford}, \& {Stern}}]{Elston06}
{Elston}, R.~J., {Gonzalez}, A.~H., {McKenzie}, E., {Brodwin}, M., {Brown},
  M.~J.~I., {Cardona}, G., {Dey}, A., {Dickinson}, M., {Eisenhardt}, P.~R.,
  {Jannuzi}, B.~T., {Lin}, Y.-T., {Mohr}, J.~J., {Raines}, S.~N., {Stanford},
  S.~A., \& {Stern}, D. 2006, \apj, 639, 816

\bibitem[{{Fiore} {et~al.}(2008){Fiore}, {Grazian}, {Santini}, {Puccetti},
  {Brusa}, {Feruglio}, {Fontana}, {Giallongo}, {Comastri}, {Gruppioni},
  {Pozzi}, {Zamorani}, \& {Vignali}}]{Fiore08}
{Fiore}, F., {Grazian}, A., {Santini}, P., {Puccetti}, S., {Brusa}, M.,
  {Feruglio}, C., {Fontana}, A., {Giallongo}, E., {Comastri}, A., {Gruppioni},
  C., {Pozzi}, F., {Zamorani}, G., \& {Vignali}, C. 2008, \apj, 672, 94

\bibitem[{{Giavalisco} {et~al.}(2004){Giavalisco}, {Ferguson}, {Koekemoer},
  {Dickinson}, {Alexander}, {Bauer}, {Bergeron}, {Biagetti}, {Brandt},
  {Casertano}, {Cesarsky}, {Chatzichristou}, {Conselice}, {Cristiani}, {Da
  Costa}, {Dahlen}, {de Mello}, {Eisenhardt}, {Erben}, {Fall}, {Fassnacht},
  {Fosbury}, {Fruchter}, {Gardner}, {Grogin}, {Hook}, {Hornschemeier}, {Idzi},
  {Jogee}, {Kretchmer}, {Laidler}, {Lee}, {Livio}, {Lucas}, {Madau},
  {Mobasher}, {Moustakas}, {Nonino}, {Padovani}, {Papovich}, {Park},
  {Ravindranath}, {Renzini}, {Richardson}, {Riess}, {Rosati}, {Schirmer},
  {Schreier}, {Somerville}, {Spinrad}, {Stern}, {Stiavelli}, {Strolger},
  {Urry}, {Vandame}, {Williams}, \& {Wolf}}]{Giavalisco04}
{Giavalisco}, M., {Ferguson}, H.~C., {Koekemoer}, A.~M., {Dickinson}, M.,
  {Alexander}, D.~M., {Bauer}, F.~E., {Bergeron}, J., {Biagetti}, C., {Brandt},
  W.~N., {Casertano}, S., {Cesarsky}, C., {Chatzichristou}, E., {Conselice},
  C., {Cristiani}, S., {Da Costa}, L., {Dahlen}, T., {de Mello}, D.,
  {Eisenhardt}, P., {Erben}, T., {Fall}, S.~M., {Fassnacht}, C., {Fosbury}, R.,
  {Fruchter}, A., {Gardner}, J.~P., {Grogin}, N., {Hook}, R.~N.,
  {Hornschemeier}, A.~E., {Idzi}, R., {Jogee}, S., {Kretchmer}, C., {Laidler},
  V., {Lee}, K.~S., {Livio}, M., {Lucas}, R., {Madau}, P., {Mobasher}, B.,
  {Moustakas}, L.~A., {Nonino}, M., {Padovani}, P., {Papovich}, C., {Park}, Y.,
  {Ravindranath}, S., {Renzini}, A., {Richardson}, M., {Riess}, A., {Rosati},
  P., {Schirmer}, M., {Schreier}, E., {Somerville}, R.~S., {Spinrad}, H.,
  {Stern}, D., {Stiavelli}, M., {Strolger}, L., {Urry}, C.~M., {Vandame}, B.,
  {Williams}, R., \& {Wolf}, C. 2004, \apjl, 600, L93

\bibitem[{{Houck} {et~al.}(2005){Houck}, {Soifer}, {Weedman}, {Higdon},
  {Higdon}, {Herter}, {Brown}, {Dey}, {Jannuzi}, {Le Floc'h}, {Rieke}, {Armus},
  {Charmandaris}, {Brandl}, \& {Teplitz}}]{Houck05}
{Houck}, J.~R., {Soifer}, B.~T., {Weedman}, D., {Higdon}, S.~J.~U., {Higdon},
  J.~L., {Herter}, T., {Brown}, M.~J.~I., {Dey}, A., {Jannuzi}, B.~T., {Le
  Floc'h}, E., {Rieke}, M., {Armus}, L., {Charmandaris}, V., {Brandl}, B.~R.,
  \& {Teplitz}, H.~I. 2005, \apjl, 622, L105

\bibitem[{{Jannuzi} \& {Dey}(1999)}]{JannuziDey99}
{Jannuzi}, B.~T., \& {Dey}, A. 1999, in Astronomical Society of the Pacific
  Conference Series, Vol. 191, Photometric Redshifts and the Detection of High
  Redshift Galaxies, ed. R.~{Weymann}, L.~{Storrie-Lombardi}, M.~{Sawicki}, \&
  R.~{Brunner}, 111--+

\bibitem[{{Le Floc'h} {et~al.}(2005){Le Floc'h}, {Papovich}, {Dole}, {Bell},
  {Lagache}, {Rieke}, {Egami}, {P{\'e}rez-Gonz{\'a}lez}, {Alonso-Herrero},
  {Rieke}, {Blaylock}, {Engelbracht}, {Gordon}, {Hines}, {Misselt}, {Morrison},
  \& {Mould}}]{LeFloch05}
{Le Floc'h}, E., {Papovich}, C., {Dole}, H., {Bell}, E.~F., {Lagache}, G.,
  {Rieke}, G.~H., {Egami}, E., {P{\'e}rez-Gonz{\'a}lez}, P.~G.,
  {Alonso-Herrero}, A., {Rieke}, M.~J., {Blaylock}, M., {Engelbracht}, C.~W.,
  {Gordon}, K.~D., {Hines}, D.~C., {Misselt}, K.~A., {Morrison}, J.~E., \&
  {Mould}, J. 2005, \apj, 632, 169

\bibitem[{{Melbourne} {et~al.}(2008){Melbourne}, {Ammons}, {Wright},
  {Metevier}, {Steinbring}, {Max}, {Koo}, {Larkin}, \& {Barczys}}]{Melbourne08}
{Melbourne}, J., {Ammons}, M., {Wright}, S.~A., {Metevier}, A., {Steinbring},
  E., {Max}, C., {Koo}, D.~C., {Larkin}, J.~E., \& {Barczys}, M. 2008, \aj,
  135, 1207

\bibitem[{{Melbourne} {et~al.}(2007){Melbourne}, {Dawson}, {Koo}, {Max},
  {Larkin}, {Wright}, {Steinbring}, {Barczys}, {Aldering}, {Barbary}, {Doi},
  {Fadeyev}, {Goldhaber}, {Hattori}, {Ihara}, {Kashikawa}, {Konishi},
  {Kowalski}, {Kuznetsova}, {Lidman}, {Morokuma}, {Perlmutter}, {Rubin},
  {Schlegel}, {Spadafora}, {Takanashi}, \& {Yasuda}}]{MelbourneSN07}
{Melbourne}, J., {Dawson}, K.~S., {Koo}, D.~C., {Max}, C., {Larkin}, J.~E.,
  {Wright}, S.~A., {Steinbring}, E., {Barczys}, M., {Aldering}, G., {Barbary},
  K., {Doi}, M., {Fadeyev}, V., {Goldhaber}, G., {Hattori}, T., {Ihara}, Y.,
  {Kashikawa}, N., {Konishi}, K., {Kowalski}, M., {Kuznetsova}, N., {Lidman},
  C., {Morokuma}, T., {Perlmutter}, S., {Rubin}, D., {Schlegel}, D.~J.,
  {Spadafora}, A.~L., {Takanashi}, N., \& {Yasuda}, N. 2007, \aj, 133, 2709

\bibitem[{{Melbourne} {et~al.}(2005){Melbourne}, {Wright}, {Barczys},
  {Bouchez}, {Chin}, {van Dam}, {Hartman}, {Johansson}, {Koo}, {Lafon},
  {Larkin}, {Le Mignant}, {Lotz}, {Max}, {Pennington}, {Stomski}, {Summers}, \&
  {Wizinowich}}]{Melbourneetal05}
{Melbourne}, J., {Wright}, S.~A., {Barczys}, M., {Bouchez}, A.~H., {Chin}, J.,
  {van Dam}, M.~A., {Hartman}, S., {Johansson}, E., {Koo}, D.~C., {Lafon}, R.,
  {Larkin}, J., {Le Mignant}, D., {Lotz}, J., {Max}, C.~E., {Pennington},
  D.~M., {Stomski}, P.~J., {Summers}, D., \& {Wizinowich}, P.~L. 2005, \apjl,
  625, L27

\bibitem[{{Monet}(1998)}]{Monet98}
{Monet}, D.~G. 1998, in Bulletin of the American Astronomical Society, Vol.~30,
  Bulletin of the American Astronomical Society, 1427--+

\bibitem[{{Peng} {et~al.}(2002){Peng}, {Ho}, {Impey}, \& {Rix}}]{Peng02}
{Peng}, C.~Y., {Ho}, L.~C., {Impey}, C.~D., \& {Rix}, H.-W. 2002, \aj, 124, 266

\bibitem[{{Pope} {et~al.}(2005){Pope}, {Borys}, {Scott}, {Conselice},
  {Dickinson}, \& {Mobasher}}]{Pope05}
{Pope}, A., {Borys}, C., {Scott}, D., {Conselice}, C., {Dickinson}, M., \&
  {Mobasher}, B. 2005, \mnras, 358, 149

\bibitem[{{Rieke} {et~al.}(2004){Rieke}, {Young}, {Engelbracht}, {Kelly},
  {Low}, {Haller}, {Beeman}, {Gordon}, {Stansberry}, {Misselt}, {Cadien},
  {Morrison}, {Rivlis}, {Latter}, {Noriega-Crespo}, {Padgett}, {Stapelfeldt},
  {Hines}, {Egami}, {Muzerolle}, {Alonso-Herrero}, {Blaylock}, {Dole}, {Hinz},
  {Le Floc'h}, {Papovich}, {P{\'e}rez-Gonz{\'a}lez}, {Smith}, {Su}, {Bennett},
  {Frayer}, {Henderson}, {Lu}, {Masci}, {Pesenson}, {Rebull}, {Rho}, {Keene},
  {Stolovy}, {Wachter}, {Wheaton}, {Werner}, \& {Richards}}]{Rieke04}
{Rieke}, G.~H., {Young}, E.~T., {Engelbracht}, C.~W., {Kelly}, D.~M., {Low},
  F.~J., {Haller}, E.~E., {Beeman}, J.~W., {Gordon}, K.~D., {Stansberry},
  J.~A., {Misselt}, K.~A., {Cadien}, J., {Morrison}, J.~E., {Rivlis}, G.,
  {Latter}, W.~B., {Noriega-Crespo}, A., {Padgett}, D.~L., {Stapelfeldt},
  K.~R., {Hines}, D.~C., {Egami}, E., {Muzerolle}, J., {Alonso-Herrero}, A.,
  {Blaylock}, M., {Dole}, H., {Hinz}, J.~L., {Le Floc'h}, E., {Papovich}, C.,
  {P{\'e}rez-Gonz{\'a}lez}, P.~G., {Smith}, P.~S., {Su}, K.~Y.~L., {Bennett},
  L., {Frayer}, D.~T., {Henderson}, D., {Lu}, N., {Masci}, F., {Pesenson}, M.,
  {Rebull}, L., {Rho}, J., {Keene}, J., {Stolovy}, S., {Wachter}, S.,
  {Wheaton}, W., {Werner}, M.~W., \& {Richards}, P.~L. 2004, \apjs, 154, 25

\bibitem[{{Sersic}(1968)}]{Sersic68}
{Sersic}, J.~L. 1968, {Atlas de galaxias australes} (Cordoba, Argentina:
  Observatorio Astronomico, 1968)

\bibitem[{{van Dam} {et~al.}(2006){van Dam}, {Bouchez}, {Le Mignant},
  {Johansson}, {Wizinowich}, {Campbell}, {Chin}, {Hartman}, {Lafon}, {Stomski},
  \& {Summers}}]{vanDam06}
{van Dam}, M.~A., {Bouchez}, A.~H., {Le Mignant}, D., {Johansson}, E.~M.,
  {Wizinowich}, P.~L., {Campbell}, R.~D., {Chin}, J.~C.~Y., {Hartman}, S.~K.,
  {Lafon}, R.~E., {Stomski}, Jr., P.~J., \& {Summers}, D.~M. 2006, \pasp, 118,
  310

\bibitem[{{Weedman} {et~al.}(2006){Weedman}, {Le Floc'h}, {Higdon}, {Higdon},
  \& {Houck}}]{Weedman06}
{Weedman}, D.~W., {Le Floc'h}, E., {Higdon}, S.~J.~U., {Higdon}, J.~L., \&
  {Houck}, J.~R. 2006, \apj, 638, 613

\bibitem[{{Werner} {et~al.}(2004){Werner}, {Roellig}, {Low}, {Rieke}, {Rieke},
  {Hoffmann}, {Young}, {Houck}, {Brandl}, {Fazio}, {Hora}, {Gehrz}, {Helou},
  {Soifer}, {Stauffer}, {Keene}, {Eisenhardt}, {Gallagher}, {Gautier}, {Irace},
  {Lawrence}, {Simmons}, {Van Cleve}, {Jura}, {Wright}, \&
  {Cruikshank}}]{Werner04}
{Werner}, M.~W., {Roellig}, T.~L., {Low}, F.~J., {Rieke}, G.~H., {Rieke}, M.,
  {Hoffmann}, W.~F., {Young}, E., {Houck}, J.~R., {Brandl}, B., {Fazio}, G.~G.,
  {Hora}, J.~L., {Gehrz}, R.~D., {Helou}, G., {Soifer}, B.~T., {Stauffer}, J.,
  {Keene}, J., {Eisenhardt}, P., {Gallagher}, D., {Gautier}, T.~N., {Irace},
  W., {Lawrence}, C.~R., {Simmons}, L., {Van Cleve}, J.~E., {Jura}, M.,
  {Wright}, E.~L., \& {Cruikshank}, D.~P. 2004, \apjs, 154, 1

\bibitem[{{Wilson} {et~al.}(2003){Wilson}, {Eikenberry}, {Henderson},
  {Hayward}, {Carson}, {Pirger}, {Barry}, {Brandl}, {Houck}, {Fitzgerald}, \&
  {Stolberg}}]{Wilson03}
{Wilson}, J.~C., {Eikenberry}, S.~S., {Henderson}, C.~P., {Hayward}, T.~L.,
  {Carson}, J.~C., {Pirger}, B., {Barry}, D.~J., {Brandl}, B.~R., {Houck},
  J.~R., {Fitzgerald}, G.~J., \& {Stolberg}, T.~M. 2003, in Presented at the
  Society of Photo-Optical Instrumentation Engineers (SPIE) Conference, Vol.
  4841, Instrument Design and Performance for Optical/Infrared Ground-based
  Telescopes. Edited by Iye, Masanori; Moorwood, Alan F. M. Proceedings of the
  SPIE, Volume 4841, pp. 451-458 (2003)., ed. M.~{Iye} \& A.~F.~M. {Moorwood},
  451--458

\bibitem[{{Wizinowich} {et~al.}(2000){Wizinowich}, {Acton}, {Shelton},
  {Stomski}, {Gathright}, {Ho}, {Lupton}, {Tsubota}, {Lai}, {Max}, {Brase},
  {An}, {Avicola}, {Olivier}, {Gavel}, {Macintosh}, {Ghez}, \&
  {Larkin}}]{Wizinowich00}
{Wizinowich}, P., {Acton}, D.~S., {Shelton}, C., {Stomski}, P., {Gathright},
  J., {Ho}, K., {Lupton}, W., {Tsubota}, K., {Lai}, O., {Max}, C., {Brase}, J.,
  {An}, J., {Avicola}, K., {Olivier}, S., {Gavel}, D., {Macintosh}, B., {Ghez},
  A., \& {Larkin}, J. 2000, \pasp, 112, 315

\bibitem[{{Wright} {et~al.}(2007){Wright}, {Larkin}, {Barczys}, {Erb},
  {Iserlohe}, {Krabbe}, {Law}, {McElwain}, {Quirrenbach}, {Steidel}, \&
  {Weiss}}]{Wright07}
{Wright}, S.~A., {Larkin}, J.~E., {Barczys}, M., {Erb}, D.~K., {Iserlohe}, C.,
  {Krabbe}, A., {Law}, D.~R., {McElwain}, M.~W., {Quirrenbach}, A., {Steidel},
  C.~C., \& {Weiss}, J. 2007, \apj, 658, 78

\bibitem[{{Zirm} {et~al.}(2007){Zirm}, {van der Wel}, {Franx}, {Labb{\'e}},
  {Trujillo}, {van Dokkum}, {Toft}, {Daddi}, {Rudnick}, {Rix},
  {R{\"o}ttgering}, \& {van der Werf}}]{Zirm07}
{Zirm}, A.~W., {van der Wel}, A., {Franx}, M., {Labb{\'e}}, I., {Trujillo}, I.,
  {van Dokkum}, P., {Toft}, S., {Daddi}, E., {Rudnick}, G., {Rix}, H.-W.,
  {R{\"o}ttgering}, H.~J.~A., \& {van der Werf}, P. 2007, \apj, 656, 66

\end{thebibliography}

\end{document}